\documentclass[
    aps,
    prx,
    superscriptaddress,
    twocolumn,
    a4paper,
    floatfix,
    longbibliography
]{revtex4-2}

\usepackage[american]{babel}

\usepackage{amsmath}
\usepackage{amssymb}
\usepackage{graphicx}
\usepackage[caption=false]{subfig}
\usepackage{color}
\usepackage[percent]{overpic}
\usepackage{bm}
\usepackage{rotating}

\usepackage{hyperref}
\usepackage{dsfont}

\usepackage{pbox}
\usepackage{array}
\usepackage{physics}
\usepackage{mathtools}
\usepackage{enumerate}
\usepackage{leftidx}
\usepackage{comment}
\usepackage{booktabs}

\usepackage{xcolor}

\usepackage[normalem]{ulem}

\usepackage{float}

\usepackage{cleveref}

\DeclareFontFamily{U}{BOONDOX-calo}{\skewchar\font=45 }
\DeclareFontShape{U}{BOONDOX-calo}{m}{n}{
  <-> s*[1.05] BOONDOX-r-calo}{}
\DeclareFontShape{U}{BOONDOX-calo}{b}{n}{
  <-> s*[1.05] BOONDOX-b-calo}{}
\DeclareMathAlphabet{\mathcalboondox}{U}{BOONDOX-calo}{m}{n}
\SetMathAlphabet{\mathcalboondox}{bold}{U}{BOONDOX-calo}{b}{n}
\DeclareMathAlphabet{\mathbcalboondox}{U}{BOONDOX-calo}{b}{n}

\newcommand{\ii}{\mathrm{i}}

\usepackage{suffix}

\usepackage[export]{adjustbox}

\newcommand{\di}{\mathrm{d}}

\DeclareFontFamily{OMS}{oasy}{\skewchar\font48 }
\DeclareFontShape{OMS}{oasy}{m}{n}{%
         <-5.5> oasy5     <5.5-6.5> oasy6
      <6.5-7.5> oasy7     <7.5-8.5> oasy8
      <8.5-9.5> oasy9     <9.5->  oasy10
      }{}
\DeclareFontShape{OMS}{oasy}{b}{n}{%
       <-6> oabsy5
      <6-8> oabsy7
      <8->  oabsy10
      }{}
\DeclareSymbolFont{oasy}{OMS}{oasy}{m}{n}
\SetSymbolFont{oasy}{bold}{OMS}{oasy}{b}{n}

\DeclareMathSymbol{\smallleftarrow}     {\mathrel}{oasy}{"20}
\DeclareMathSymbol{\smallrightarrow}    {\mathrel}{oasy}{"21}
\DeclareMathSymbol{\smallleftrightarrow}{\mathrel}{oasy}{"24}

\DeclareMathAlphabet\mathbfcal{OMS}{cmsy}{b}{n}

\begin{document}

    \title{Entanglement harvesting in the presence of cavities}


    \author{Jannik Str\"ohle}
    \email{jannik.stroehle@uni-ulm.de}
	\affiliation{Institut f{\"u}r Quantenphysik and Center for Integrated Quantum
    Science and Technology (IQST), Universit{\"a}t Ulm, Albert-Einstein-Allee 11, D-89069 Ulm, Germany}

    \author{Nikolija Mom\v{c}ilovi\'{c}}
	\affiliation{Institut f{\"u}r Quantenphysik and Center for Integrated Quantum
    Science and Technology (IQST), Universit{\"a}t Ulm, Albert-Einstein-Allee 11, D-89069 Ulm, Germany}

\begin{abstract}
So far, entanglement harvesting has been extensively studied in free space setups. Here, we provide a detailed analytical and numerical analysis of entanglement harvesting in cavities. Specifically, we adiabatically couple the quantized electromagnetic field to two identical Gaussian detectors located on the symmetry axis of a cylindrical cavity. Our numerical investigations reveal a strong dependence on the cavity length, while showing invariance under changes in the cavity radius in regimes of maximal entanglement. 
Moreover, we identify different scalings of the detector system parameters for entanglement inside and outside the light cone. Finally, we uncover a strong dependence of the harvested correlations on the cavity induced parity of the electromagnetic field.
\end{abstract}

\maketitle

\section{Introduction}

It is well established that two detectors coupling solely to the quantum vacuum can become entangled over time~\cite{Summers1,Summers2}. The pioneering work on quantifying and accessing this phenomenon—now known as entanglement harvesting—was carried out by \mbox{\citeauthor{valentini1991non}}~\cite{valentini1991non} and \mbox{\citeauthor{reznik2005violating}}~\cite{reznik2005violating,reznik2003entanglement}. In the literature, entanglement harvesting is most commonly investigated using the Unruh–DeWitt (UDW) model, which describes a spherically symmetric two-level quantum system interacting with a massless scalar quantum field~\cite{pozas2015harvesting,martin2016spacetime,gale2023relativistic,stritzelberger2021entanglement}. Despite its simplicity, the UDW model captures essential features of light–matter interaction and has inspired several experimental proposals, including implementations based on superconducting circuits~\cite{sabin2012extracting} and accelerated atoms in the context of the Unruh effect~\cite{Benatti,Salton,LiuEntanglement,ZhangYu}.
However, due to the scalar field and linear coupling, the UDW model does not take into account the vector nature of the electromagnetic field. This is overcome by \mbox{\citeauthor{pozas2016entanglement}}, where entanglement harvesting based on the dipole-field coupling between a hydrogen-like atom and the electromagnetic vacuum field is studied, thereby incorporating effects of anisotropies and orientational dependencies of the atomic structure~\cite{pozas2016entanglement}.\\
Despite this improvement, entanglement harvesting is so far considered almost exclusively in free space setups~\cite{pozas2015harvesting,pozas2016entanglement,BuhmannEntanglement,Foo,Gallock,Naeem}. Consequently, modifications well established in cavity QED and highly restricted in free space, such as Purcell enhancement~\cite{Reiserer,BlaisCirc}, geometrical control of correlations~\cite{Joulain,Caze} and spatial mode filtering~\cite{Kleppner} are omitted. 
The advantages of exploiting cavity mediated setups to investigate non classical correlations have already been demonstrated in fields closely related to entanglement harvesting, where multiple efforts have been made in the last decades. These include~\citeauthor{HarocheRyd1}, establishing the toolbox for probing cavity vacuum correlations by measuring and controlling atom-photon entanglement, via Rydberg atoms in superconducting microwave cavities~\cite{HarocheRyd1,HarocheRyd2}. This was followed by the efforts of\mbox{~\citeauthor{Reiserer}} coupling single atoms trapped in resonators, enabling novel techniques for quantum information exchange and entanglement distribution~\cite{Reiserer2}. Moreover, \mbox{\citeauthor{Kastor}} proposed a scheme to maximally entangle two atoms in an optical cavity via a novel technique using dissipative state preparation enabled by cavity decay~\cite{Kastor}. 
Additional methods involve the concept of entanglement farming~\cite{MartinFarming}, using repeated weak interactions of probe pairs with an optical cavity, entanglement enhancing of two accelerated detectors via cavity boundaries~\cite{Liu,Barman}, Einstein-Podolsky-Rosen (EPR)-entangled light pulses from a single trapped atom in a high finesse cavity~\cite{Morigi} and the generation of maximally entangled states in a large detuned atom-cavity setup~\cite{Shi}. 
On top of that comes the rapidly advancing topic of exploiting different cavity environments and other objects like gratings and lasers to push the amplitude and lifetime of entanglement to its maximum. This includes cavity based honeycomb lattices~\cite{Arreyes}, double layer graphene setups in a microcavity~\cite{Ardenghi}, coupled cavity arrays~\cite{Liew} and squeezed-reservoir engineering in a cavity~\cite{Yang}. 
While these experiments confirm the advantages of cavities in quantum information, it is essential to investigate the extent to which quantum correlations can be generated and amplified in cavity environments. Thus, and in contrast to prior literature, we will directly account for different cavity regimes ranging from microcavities to optical cavities and including also waveguides and disc cavities. 
Focusing on entanglement harvesting, we solely focus on entanglement produced by the quantum vacuum, isolating the intrinsic cavity vacuum correlations as the source of entanglement, thus bridging the gap between entanglement harvesting and the rich topic of cavity mediated quantum correlations. \\
Since we consider the whole vector nature of the cavity field, the influence of longitudinal and transversal mode degrees of freedom as possible effects on entanglement caused by the fields parity are taken into account. \\
Other features not available in free space such as geometrical field amplitudes and control of
the field’s mode density are investigated and exploited to maximize entanglement.
Smooth switchings help to find optimal interaction windows based on the cavity regime as also on the detector parameters. This provides the missing link between entanglement harvesting in different cavity regimes as also with the free space regime.\\  
This paper is organized as follows: In Sec.~\ref{Basics} the formalism for entanglement harvesting inside a general cavity for two general two-level systems is established. In Sec.~\ref{NegCal} we particularize this setup to two Gaussian detectors placed opposite on the symmetry axis of a cylindrical cavity. In Sec.~\ref{NegCalB} we investigate the therewith obtained results numerically. Therefore, different cavity lengths and radii for two sets of different detector parameters are considered. Afterwards the detector parameters are varied, while comparing four different cavity regimes which are the micro cavity, the waveguide, the disc cavity and the optical cavity.  
\label{Intro}

\section{Entaglement harversting in the presence of an electromagnetic field}
\label{Basics}

Consider two detectors, Alice located at $\boldsymbol{r}_A$  and Bob at $\boldsymbol{r}_B$, separated by a distance \mbox{$\abs{\boldsymbol{r}_A-\boldsymbol{r}_B}$} and coupled to the vacuum of an electromagnetic field. Provided each party is carrying an electric charge $e$, both parties can get entangled with the help of classical communication, e.g., by exchanging a photon~\cite{nielsen2010quantum}. 
Even though the field vacuum has no photons, the vacuum undergoes quantum fluctuations. The quantum nature of these fluctuations can generate correlations which do not exist in classical communication such as entanglement of spacelike separated points~\cite{reznik2005violating}.
Throughout the article, spacelike separation refers to detectors whose interaction regions are spacelike separated, precluding any causal influence. In this case, any correlations are mediated exclusively by vacuum correlations of the field, corresponding to vanishing field commutator at two spacelike separated coordinates. Timelike separation instead, refers to the regime where entanglement between the two detectors can be created by both,  pre-existing vacuum correlations but also signaling (communication) between the two detectors. Here the field commutator does not necessarily need to vanish.\\
Thus, while classical communication is limited by the speed of light, the quantum vacuum allows even for correlations of two detectors separated by \mbox{$\mathrm{d}s^2=c^2\mathrm{d}t^2-\mathrm{d}x^2<0$}, such that the detectors interact for a time \mbox{$t<\abs{\boldsymbol{r}_A-\boldsymbol{r}_B}/c$}.
The quantification of the thereby generated entanglement is known in the literature as entanglement harvesting~\cite{pozas2015harvesting,pozas2016entanglement}. Although it is possible to tune the parameters of the system of interest such that only spacelike entanglement is measured, entanglement harvesting itself accounts for measuring entanglement in the spacelike as well as in the timelike regime. 
\\
To model the interaction between the two detectors and the cavity field, we assume each detector as an effective one particle system. Each particle is modeled by a classical nucleus much heavier than the electron. Besides describing each particle as a single point like dipole in space we directly account for the spatial extension of each particle by introducing a smearing. Thus, we define the dipole operator 
\begin{equation}
    \boldsymbol{\hat{d}}(t,\boldsymbol{r})=e\left(e^{i\Omega t}\boldsymbol{F}(\boldsymbol{r})\hat{\sigma}_{+}+ e^{-i\Omega t}\boldsymbol{F}^\dagger(\boldsymbol{r})\hat{\sigma}_{-} \right),
\end{equation}
where 
\begin{equation}
    \boldsymbol{F}(\boldsymbol{r})=\Psi_{e}^*(\boldsymbol{r})\boldsymbol{r}\Psi_{g}(\boldsymbol{r})
    \label{Smear1}
\end{equation}
describes the smearing function containing the detectors wave functions \mbox{$\Psi_{g} (\bm r)$} and \mbox{$\Psi_{e}(\bm r)$} for ground and excited state, respectively.
The dynamics of the total system is then governed in the interaction picture by the Hamiltonian~\cite{Lopp2,strohle2024dimensional}, 
\begin{equation}
    \hat{H}_I (t) =\sum_{\alpha=A,B}\chi_{\alpha}(t)\int\mathrm{d}^3\boldsymbol{x}\,{\boldsymbol{\hat{d}}_{\alpha}(t,\boldsymbol{x}-\boldsymbol{x}_{\alpha})}{\boldsymbol{\hat{E}}(t,\boldsymbol{x})},
    \label{Eq:IH}
\end{equation}
 and the electric field
\begin{equation}
        \hat{\boldsymbol{E}}(\boldsymbol{x},t)=\sum_{\boldsymbol{j},\mu}\left[A_{\boldsymbol{j},\mu}\hat{a}_{\boldsymbol{j},\mu}(t)\boldsymbol{u}_{\boldsymbol{j},\mu}(\boldsymbol{x})+\mathrm{h.c.}\right],
        \label{EField}
\end{equation}
with the complex amplitude $A_{\boldsymbol{j},\mu}=i\sqrt{\frac{\hbar\omega_{\boldsymbol{j},\mu}}{2\epsilon_0}}$ and the sum over the modes $\boldsymbol{j}=(n,m,l)$ and polarizations $\mu=(\mu_1,\mu_2)$. To obtain adiabatic interaction between the field and detector we choose Gaussian switching
\begin{align}
    \chi_\alpha(t) =\exp \left( - \frac{(t-t_\alpha)^2}{2 T^2} \right),
    \label{GS}
\end{align}
where $t_\alpha$ is the proper time of detector $\alpha$.\\
To quantify entanglement between two detectors, we define the negativity
\begin{equation}
    \mathcal{N}=\max\{0,\mathcal{N}^{(2)}\},
\end{equation}
as entanglement measure.
The negativity estimator $\mathcal{N}^{(2)}$ for two identical detectors switched on for a time $T$ up to second order Dyson reduces to the expression \footnote{This simplified expression is based on the assumption that the local correlations for two identical detectors are identical. This assumption holds inside a cylindrical cavity if, in addition, a symmetric arrangement of the detectors is provided.}:  
\begin{equation}
    \mathcal{N}^{(2)}= \abs{\mathcal{M}}-\mathcal{L},
    \label{eq4.29}
\end{equation}
where we define local correlations $\mathcal{L}$ and non-local correlations $|\mathcal{M}|$.
Local correlations arise from each detectors self-noise, i.e. the noise each detector picks up from its own (local) coupling to the field along its worldline. In contrast, non-local correlations, are correlations between the two quantum detectors arising from the coupling to the quantum vacuum, which itself is correlated across all points in space and time (non-local) within the cavity. Although the detectors only interact locally with the vacuum, the correlations appear to be non-local. Thus, the entanglement between the two detectors, as well as the entanglement each detector accumulates locally with the quantum vacuum, is quantified by the non-local correlations. 
Accordingly, the negativity can only take values different from zero when the non-local correlations $\mathcal{M}$ exceed the strictly positive local correlations
\mbox{$\mathcal{L}$}. While former are given by the off-diagonal elements of the time evolved density matrix, the latter resemble the respective block diagonal elements, reading ~\cite{pozas2016entanglement}:
\begin{widetext}
\begin{subequations}
\begin{align}
     \mathcal{L}=&\hspace{.4cm}e^2\int_{-\infty}^{\infty}\mathrm{d}t_1\int_{-\infty}^{\infty}\mathrm{d}t_2\int\mathrm{d}^3 x_1\int\mathrm{d}^3 x_2\,e^{i\Omega(t_1-t_2)}\chi_{\alpha}(t_1)\chi_{\alpha}(t_2) \boldsymbol{F}^{T}(\boldsymbol{x}_2-\boldsymbol{x}_{\alpha})\boldsymbol{W}(\boldsymbol{x}_2,\boldsymbol{x}_1,t_2,t_1)\boldsymbol{F}(\boldsymbol{x}_1-\boldsymbol{x}_{\alpha}),
    \label{eq4.22}\\
\begin{split}
    \mathcal{M}=&-e^2\int_{-\infty}^{\infty}\mathrm{d}t_1\int_{-\infty}^{t_1}\mathrm{d}t_2\int\mathrm{d}^3 x_1\int\mathrm{d}^3 x_2\,e^{i\Omega(t_1+t_2)} [\chi_{A}(t_1)\chi_{B}(t_2)\boldsymbol{F}^{T}(\boldsymbol{x}_1-\boldsymbol{x}_{A})\boldsymbol{W}(\boldsymbol{x}_1,\boldsymbol{x}_2,t_1,t_2)\boldsymbol{F}_{\mathcal{R}}(\boldsymbol{x}_2-\boldsymbol{x}_{B})\\
    &+\chi_{B}(t_1)\chi_{A}(t_2)\boldsymbol{F}_{\mathcal{R}}^{T}(\boldsymbol{x}_1-\boldsymbol{x}_{B})\boldsymbol{W}(\boldsymbol{x}_1,\boldsymbol{x}_2,t_1,t_2)\boldsymbol{F}(\boldsymbol{x}_2-\boldsymbol{x}_{A})],
\end{split}
    \label{eq4.23}
\end{align}
\label{eq4}%
\end{subequations}
\end{widetext}
with the Wightman function
\begin{equation}
\begin{aligned}
    \boldsymbol{W}(\boldsymbol{x}_2,\boldsymbol{x}_1,t_2,t_1)&=\bra{0}\hat{ \bm E}(\boldsymbol{x}_2,t_2) \,\hat{\bm E}^\dagger (\boldsymbol{x}_1,t_1)\ket{0},
\end{aligned}
      \label{eq4.24}
\end{equation}
resembling a correlation function of the electric field's vacuum evaluated at different points in space and time.
Note that in~\eqref{eq4.22} a symmetric arrangement of the two detectors is required, leaving the local correlations invariant when exchanging $\bm x_A$ with $\bm x_B$ and vice versa. Since the local correlations are not affected by time ordering and both detectors experience identical switchings, the local correlations are independent of the index $\alpha$.
Additionally, for the non-local correlations in Eq.~\eqref{eq4.23}, the smearing function \mbox{$\bm F_{\mathcal{R}}(\bm x)$} adds the relative rotation between detector $A$ and $B$, yielding a smearing different from \mbox{$\bm F(\bm x)$} for non rotation invariant wave function.  
For a detailed discussion of the negativity of this specific setup and a derivation of the negativity related quantities defined here we refer to~\cite{pozas2015harvesting,pozas2016entanglement}.

\section{Analytical Negativity}
\label{NegCal}

\begin{figure}
\includegraphics[width=7cm]{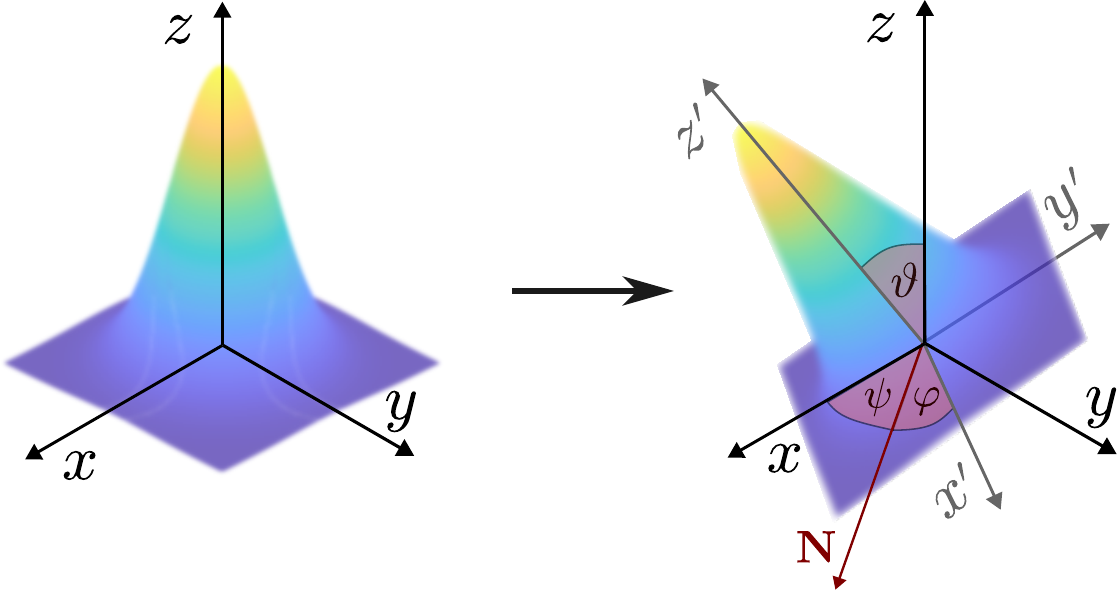}
\caption{Example of the rotation of a Gaussian state in position representation. For the Euler angles we have $\vartheta$ defining rotations from $z$ to $z'$, $\psi$ defining the rotation from $x$ to \textbf{N} and $\varphi$ for rotations from \textbf{N} to $x'$. Here \textbf{N} is the line of nodes where the $x$, $y$ surfaces of primed and unprimed coordinate system intersect. With these 3 angles all possible rotations between two coordinate systems can be described. }
\label{FigRot}
\end{figure}

We consider two  Gaussian detectors $A$ and $B$, each with one ground and one excited state with respect to the  symmetry axis of the cavity, this is the $z$-coordinate. As a result, the wave functions of the ground and excited state for the detectors $A$ and $B$ in cylindrical coordinates, i.e., $\bm r = (\rho, \phi, z)$, read~\cite{strohle2024dimensional}:
\begin{subequations}
\begin{align}
    \Psi_g^{A,B}(\boldsymbol{r}) &=\frac{e^{-(\rho^2+z^2)/(2\sigma^2)}}{\pi^{3/4}\sigma^{3/2}},
    \label{eq4.3}\\
    \Psi_e^{A,B}(\boldsymbol{r}) &= \frac{e^{-(\rho^2+z^2)/(2\sigma^2)}}{\pi^{3/4}\sigma^{3/2}}\frac{\sqrt{2}z}{\sigma},
    \label{eq4.4}
\end{align}
\label{WaveEq}%
\end{subequations}
where $\sigma$ defines the width of the wave function with transition frequency $\Omega$ between ground state and excited state. In contrast to the ground state wave function, the excited state wave function is not invariant under rotations. As a consequence, the two detectors will have a relative orientation towards each other. 
In order to incorporate this characteristic properly, the relative orientation of detector $B$ has to be expressed with respect to the reference frame of detector $A$ or vice versa. Using Euler angles $(\psi,\vartheta,\varphi)$, depicted in Fig.~\ref{FigRot}, we obtain two smearing functions
\begin{subequations}
\begin{equation}
    \boldsymbol{F}(\boldsymbol{x})=\frac{\sqrt{2}}{\pi^{3/2}\sigma^4}e^{-\rho^2/\sigma^2}e^{-z^2/\sigma^2}z\left(\begin{array}{c}
                \rho\cos(\phi)\\
                \rho\sin(\phi)	\\
                z
\end{array}\right)
    \label{eq4.12}
\end{equation}
and when transforming to the rotated frame
\begin{align}
\begin{split}
     \boldsymbol{F}_\mathcal{R}(\boldsymbol{x})=\frac{\sqrt{2}}{\pi^{3/2}\sigma^4}e^{-\rho^2/\sigma^2}e^{-z^2/\sigma^2}\left(\begin{array}{c}
                \rho\cos(\phi)\\
                \rho\sin(\phi)	\\
                z
\end{array}\right)\times\\
[\rho\sin(\vartheta)\cos(\phi+\psi)+z\cos(\vartheta)].\label{Smear2}
\end{split}
\end{align}
\label{Smear}%
\end{subequations}
A detailed calculation regarding the detector orientation is provided in App.~\ref{AppDet}.\\ 
In the following the negativity will be specified for the case of two identical Gaussian detectors placed opposite on the symmetry axis of a cylindrical cavity. 
Due to the two detectors located on the symmetry axis of a cylindrical cavity the interaction of the transversal electric modes vanishes, leaving only field modes of transversal magnetic polarizations $\mu_2$~\cite{strohle2024dimensional}. Additionally the $\varphi$-symmetry of the problem gives only non vanishing overlap between field and detectors if the polar mode number is set to zero, i.e., $n = 0$. Thus, we will without loss of generality omit the polarization and polar mode number dependence in the following. For a more compact notation we therefore rearrange the mode index to \mbox{$\bm j = (m,l)$}, containing only two indices namely the radial mode number $m$ and the longitudinal mode number $l$.\\ 
Since the orientation of the two detectors relative to each other is irrelevant when only a single detector is considered (local correlations), we use in Eq.~\eqref{eq4.22}, without loss of generality, the smearing function~\eqref{eq4.12}.
Thus, the respective detector orientation, comes only into play when calculating the non-local correlations of Eq.~\eqref{eq4.23}, by the simultaneous appearance of both smearings in Eq.~\eqref{Smear}.\\
Taking into account the Euler angles we therefore obtain for the non-local correlations
\begin{subequations}
\begin{align}
\begin{split}
    \mathcal{M}= \mathcal{A} &\frac{\cos(\vartheta)}{2} \sum_{\bm j}\xi_{\bm j} \left( E(\omega_{\bm j}, t_{BA}) + E(\omega_{\bm j}, -t_{BA}) \right)\\   &\times e^{-(\omega_{\bm j}^2 + \Omega^2 ) T^2/2} \cos(\gamma^{(-)}_l) \cos(\gamma^{(+)}_l),
    \end{split}
    \label{MExplicit}
\end{align}
and for the local correlations for each of of the two identical detectors
\begin{align}
        \mathcal{L}
        =&\mathcal{A}\sum_{\bm j}\xi_{\bm j} e^{-(\omega_{\bm j} + \Omega )^2 T^2/2}\cos^2\left(\gamma^{(-)}_l\right),
        \label{LExplicit}
\end{align}
\label{NLExplicit}%
\end{subequations}
with the prefactor of dimension length
\begin{subequations} 
    \begin{equation}
        \mathcal{A}= \frac{c \,e^2  \, T^2 \sigma^2}{2 \epsilon_0 \hbar L R^2}
        \label{Eq.A}
    \end{equation}
    and the mode dependent coefficient
    \begin{equation}
         \xi_{\bm j}=\frac{\abs{\boldsymbol{k}_{m}}^2}{J_1^2(\chi_{m})\abs{\boldsymbol{k}_{\bm j}}}e^{-\left(\abs{\boldsymbol{k}_{\bm j}}\sigma\right)^2/2},
         \label{Eq.Xi}
    \end{equation}
    where $J_n$ is the Bessel function of first kind.
    Furthermore, the phase 
    \begin{equation}
        \gamma^{(\pm)}_l\coloneqq\frac{\pi l}{2}\left(1\pm\frac{D}{L}\right),
    \end{equation}
    with the distance between the two detectors $D$, carries a minus sign to denote detector $A$'s location and a plus sign for detector $B$'s location. Additionally we defined in Eq.~\eqref{MExplicit} the mode number dependent amplitudes
    \begin{align}
         E(\omega_{\bm j},\pm t_{BA}) &=e^{\pm i\omega_{\bm j}t_{BA}}\mathrm{erfc}\left(\frac{\pm t_{BA}+iT^2\omega_{\bm j}}{\sqrt{2}T}\right),
         \label{Eq.E}
    \end{align}
    \label{Eq.Ampl}%
\end{subequations}
which is proportional to the complementary error function, $\mathrm{erfc}$ and involves the interaction time $T$ and the time delay $t_{BA}=t_B-t_A$ between the switchings. For the wave numbers in Eqs.~\eqref{Eq.Ampl} we define
\begin{align}
    |\bm k_{\bm j}|^2 = |\bm k_{m}|^2 + k_l^2 = \left(\frac{\chi_m}{R}\right)^2 + \left(\frac{l \pi}{L}\right)^2\!\!,
    \label{KS}
\end{align}
where $\chi_m$ is the $m$-th zero of $J_0$.
Detailed derivations of Eq.~\eqref{LExplicit} are laid out in App.~\ref{App1}.\\
From our results in Eqs.\,\eqref{NLExplicit} we thus obtain 
\begin{align}
\begin{split}
       \mathcal{N}^{(2)} \!= \! \mathcal{A}  &\left\{ \frac{\cos(\vartheta) }{2} \left|  \sum_{\bm j}\xi_{\bm j} \left( E(\omega_{\bm j}, t_{BA}) + E(\omega_{\bm j}, -t_{BA}) \right) \right. \right. \\
       &\left. \cdot \exp\left(-\frac{\left(\omega_{\bm j}^2 + \Omega^2 \right) T^2}{2}\right) \cos(\gamma^{(-)}_l) \cos(\gamma^{(+)}_l) \right| \\
        &\left.- \sum_{\bm j}\xi_{\bm j} \exp\left(-\frac{\left(\omega_{\bm j} + \Omega\right)^2 T^2}{2}\right) \cos^2\left(\gamma^{(-)}_l\right) \right\}    
\end{split}
    \label{eq4.31}
\end{align}
for the negativity estimator. This estimator depends on the chosen interaction given by the switchings~\eqref{GS}, the placement of each detector and the wave function of each detector  entering via the smearing function~\eqref{Smear}. 
Before starting with the numerical investigation in the following section we will give a brief classification of the underlying model. 
There are various physical platforms for the two detectors, for example two Bose-Einstein condensates (BEC) trapped in a harmonic potential. Here, the width $\sigma$ of each detector is given by the harmonic oscillator length, typically of the order of \mbox{$\sigma \approx 1 \, \text{\textmu} \mathrm{m}$}. The transition frequency $\Omega$ of the harmonic trap states, given by the harmonic oscillator length, then ranges for the most common BEC species, such as $^{87} \mathrm{Rb}$ or $^{23}\mathrm{Na}$, between ten and hundreds of Hertz~\cite{Dalfovo}.
In contrast, we find when using superconducting qubits instead, $\sigma \approx 10 \, \text{\textmu} \mathrm{m}$, but typical energy gaps of the order of a few $10 \, \mathrm{GHz}$~\cite{McKay}. These qubits can be modeled for different sizes even reaching $\sigma \geq100\, \text{\textmu} \mathrm{m}$ with $\Omega$ still in the microwave regime~\cite{Gou}.     
Another implementation involves Rydberg pairs as detectors. Here the most prominent example are circular Rydberg atoms with a transition frequency of $51.099\, \mathrm{GHz}$ and the spatial extent of the Rydberg orbits of $\sigma \approx 0.1 \, \text{\textmu} \mathrm{m}$~\cite{HarocheRyd2}.\\
With a suitable detector setup at hand, the geometrical properties of the cavity have to be chosen. These, imprinted on the correlations via the electric field~\eqref{EField}, enter by the vectorized detector field interaction. Therefore, we find as dimensionless quantities in Eq.~\eqref{eq4.31} the detector interaction time $t_{BA}/T$, the detector separation distance $D/\sigma$, the detector energy $\Omega T$ and the cavity length $L/\sigma$ and the cavities radius $R/\sigma$. While all of these quantities can have unique influences on entanglement harvesting, we will lay special focus on the cavity parameters and their influence on the detector, switching and interaction parameters in the following.

\section{Numerical Negativity}
\label{NegCalB}

\begin{figure}
\includegraphics[width=8.25cm]{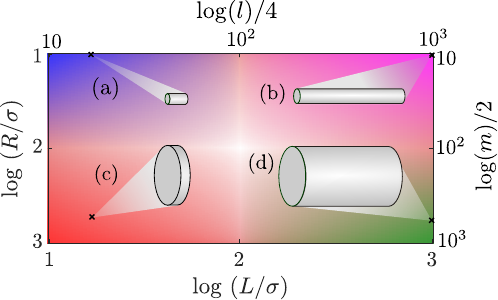}
\caption{The four cavity regimes, including the microcavity regime (blue), the waveguide regime (purple), the disc cavity regime (red) and the optical cavity regime (green). The four cases \mbox{(a)-(d)} chosen in the discussion of the detector system parameters in Figs.~\ref{Fig2} and~\ref{Fig3} are sketched and marked with crosses. The dimensionless cavity radius \mbox{$R/\sigma$} and dimensionless cavity length \mbox{$L/\sigma$} are scaled on the left and lower axis. The maximal number of modes necessary for convergence is scaled on the upper axis and the right axis for $l$ and $m$, respectively. For the smallest cavity length \mbox{$L/\sigma = 20$}, already longitudinal modes with \mbox{$\max(l) = 10^2$} suffice, while for the maximum \mbox{$L/\sigma = 10^3$} longitudinal modes up to \mbox{$\max(l) = 4\cdot 10^3$} have to be chosen to guarantee convergence. For the cavity radii stronger convergence is obtained with transversal modes up to $\max(m) = 10$ sufficient for \mbox{$R/\sigma = 5$}, while \mbox{$\max(m) = 10^3$} is needed for \mbox{$R/\sigma = 500$}.}
\label{FigReg}
\end{figure}

In the following we will investigate the negativity estimator~\eqref{eq4.31} of the two Gaussian wave packets located on the symmetry axis of the cylindrical cavity numerically. Since the respective orientation of the two detectors arises solely by a prefactor of \mbox{$\cos(\vartheta)$} in the non-local correlations, we maximize $|\mathcal{M}|$ by choosing $\cos(\vartheta) = 1$, indicated by two equally orientated detectors.\\
To briefly outline the convergence of the mode number summation in Eq.~\eqref{eq4.31} note that due to the harmonic oscillator eigenstates used for the detectors in Eq.~\eqref{WaveEq}, Gaussian convergence dictated by the dimensionless cavity lengths and cavity radii dominates. These Gaussian functions act especially for small cavity length and small cavity radii like a low pass transformation of the systems amplitudes. Thus, the maximum in supported cavity frequencies increases when enlarging the cavity.
The different cavity regimes considered and the convergence in mode numbers is shown in Fig.~\ref{FigReg}.
To differ between effects caused by changing the cavity radii and effects stemming from varying the cavity length, modes with constant $m$ but different $l$ will be called longitudinal modes while modes with different $m$ but constant $l$ will be called transversal modes. We start the numerical investigation by focusing on different geometrical regimes of the cavity, ranging from microcavity to optical cavity.   

\begin{figure*}
\includegraphics[width=17.5cm]{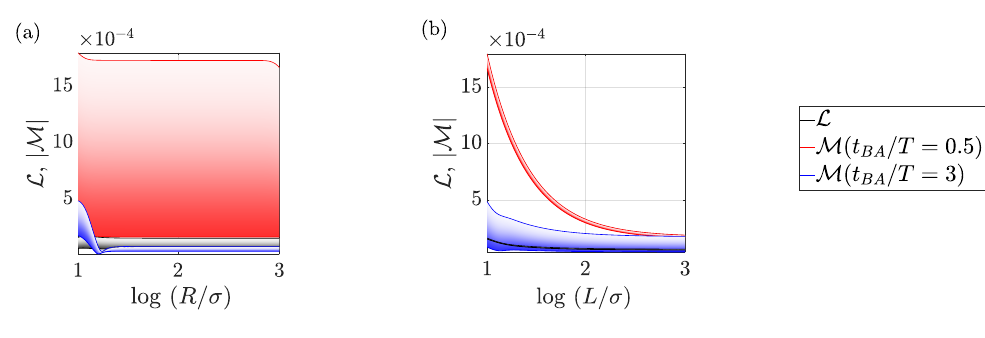}
\caption{Variation of local correlations $\mathcal{L}$ and non-local correlations $|\mathcal{M}|$ for $D/\sigma = 5$, $\Omega T = 1$ and different detector interaction times in the timelike $t_{BA}/T = 3$ and spacelike regime $t_{BA}/T = 0.5$. 
In (a) we plot $R/\sigma$ on the $x$-axis while the color gradient scales with $L/\sigma$. The upper and lower curve for each $\mathcal{L}$ and $|\mathcal{M}|$ are for the minimal cavity length $L/\sigma= 10$ and the maximal length $L/\sigma = 10^3$, respectively. The color gradient in between gives the stability of the respective correlations, i.e., for large opacity strong variation in $L/\sigma$ gives little variation in correlations and vice versa for small opacity. 
In (b) we scale $R/\sigma$ in the interval from $R/\sigma = 5$ to $R/\sigma = 10^3$ on the $x$-axis and the color gradient is scaled with $L/\sigma$ in the same interval.}
\label{Fig1Ext}
\end{figure*}

\begin{figure*}
\includegraphics[width=17.5cm]{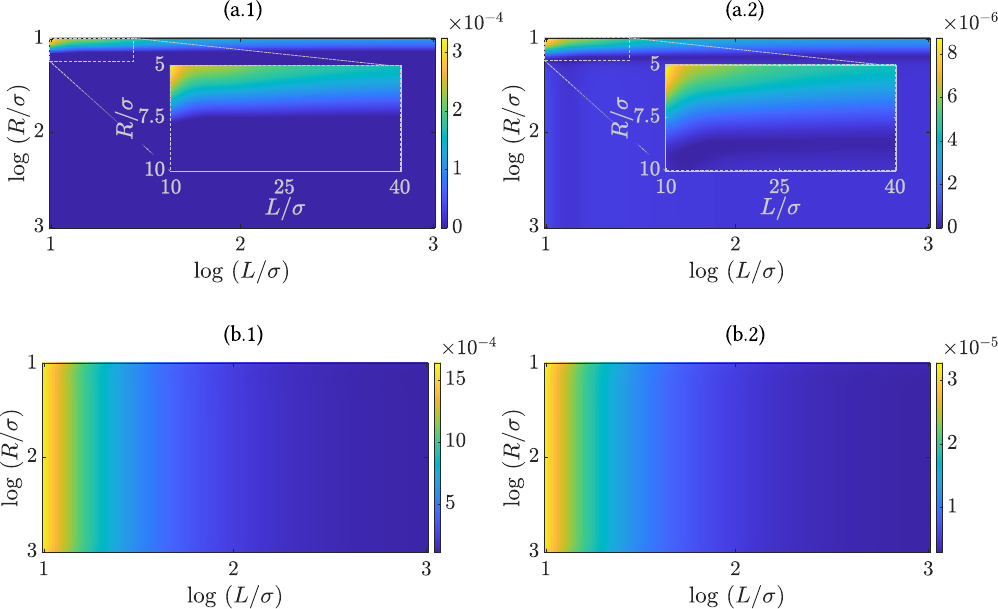}
\caption{Negativity estimator $\mathcal{N}$ for $D/ \sigma = 5$ in the timelike regime, $t_{BA}/T = 3$ (a), and the spacelike regime, $t_{BA}/T = 0.5$ (b). We choose different detector energies with $\Omega T= 1$ (a.1,  b.1) and $\Omega T = 3$ in (a.2, b.2). To cover all the cavity regimes logarithmic axis are chosen. Additionally we include in (a) a zoomed plot showing solely the microcavity regime.}
\label{Fig1}
\end{figure*}

\subsection{Imprint of the Cavity Geometry}
\label{ImprintCavGeom}

To determine the influence of the cavity geometry on the detectors entanglement we vary the dimensionless cavity length $L/\sigma$ and the dimensionless cavity radius $R/\sigma$. Since the detector wave functions have spatial extent itself we define two small offsets $\delta L/\sigma = 10$ and $\delta R/\sigma = 5$ to ensure that the detectors wave function can reach small values before colliding with the cavity walls. To investigate a broad range of cavity regimes the upper limit is chosen as \mbox{$\max(L/\sigma) = \max(R/\sigma) = 10^3$}. These regimes include the microcavity, the waveguide regime, the disc cavity and the optical cavity regime, as it is shown in Fig.~\ref{FigReg}. When further increasing $R/\sigma$ and $L/\sigma$ in the optical cavity regime the free space regime is reached. \\
We compare the scaling in cavity length of local and non-local correlations in both time and spacelike regime for different cavity radii in Fig.~\ref{Fig1Ext}(a). In Fig.~\ref{Fig1Ext}(b) we plot the same setup but for different cavity lengths and scale with cavity radius.\\
The negativity is plotted for the spacelike regime in Fig~\ref{Fig1}(a) and the timelike regime in Fig~\ref{Fig1}(b), including all four different cavity regimes.\\
Comparing the increase in cavity length with the increase in cavity radius different patterns in negativity emerge. Thus, we split the following discussion in two parts, starting with the cavity radius.

\subsubsection{Imprint of cavity radius}
    To investigate the correlations with respect to different cavity radii it is suitable to hold the longitudinal mode numbers $l$ in the summation of Eqs.~\eqref{NLExplicit} fixed. When neglecting all parts which do not scale with mode number $m$ and $R/\sigma$ the following functions mapping the whole radial scaling of the different correlations are obtained
    \begin{subequations}
        \begin{align}
            M_{l}  = &\left| \sum_m  \frac{\xi_{\bm j}}{R^2}  e^{-\omega_{\bm j}^2T^2/2} \vphantom{\frac{1}{1}}\left( E(\omega_{\bm j}, t_{BA}) \! + \! E(\omega_{\bm j},-t_{BA}) \right) \right|,
            \label{ML}\\
            L_l =  &\sum_m  \frac{\xi_{\bm j}}{R^2}  e^{-\omega_{\bm j}^2T^2/2} \lambda_{\bm j},
            \label{LL}
        \end{align}       
        \label{MLll}%
    \end{subequations}
    with the coupling of cavity frequency and detector energy
            \begin{align}
            \lambda_{\bm j} = e^{- \omega_{\bm j}\,  \Omega T^2}.
            \label{Coupllambda}
        \end{align}
The dominating term in Eqs.~\ref{MLll} -- given by the harmonic oscillator eigenstates chosen in Eq.~\eqref{WaveEq} and the Gaussian switching in Eq.~\eqref{GS} -- is a Gaussian window  acting like a low pass for each field mode, i.e., weighting small $|\bm k_{\bm j}|$ with the largest amplitude in both local and non-local correlations.
Since the Gaussian window of contributing traversal modes grows with $~\chi_m \sim R/\sigma$, we find the local correlations in Fig.~\ref{Fig1Ext}(a) invariant of $R/\sigma$. 
This is caused by the Bessel zeros in Eq.~\eqref{Eq.Xi}~\footnote{Note that \mbox{$J_1(\chi_m)^{-2} \sim \chi_m$}.}, balancing the $(R/\sigma)$ dependence of Eq.~\eqref{LL}, as we find for both, local and non local correlations in Fig.~\ref{Fig1Ext}. Here, when increasing $R/\sigma$ -- besides the geometrical amplitude $\sim R^{-2}$ truncating the amplitude of each mode -- a denser transverse mode spectrum occurs yielding constructive interference of more modes, with non neglectable contribution to the detector-field interaction. 
While for the non local correlations the constructive interference arises due to the positive nature of the single noise terms in the sum of Eq.~\eqref{LL}, for the non-local correlations in Eq.~\eqref{ML}, the amplitude \mbox{$E(\omega_{\bm j}, t_{BA}) + E(\omega_{\bm j},-t_{BA})$}, given by Eq.~\eqref{Eq.E}, dictates interference between the different transverse mode contributions.
Here, radial scaling is dominated by a Gaussian decay in \mbox{$R/\sigma$} by the \mbox{$\mathrm{erfc}$}-terms which depend strongly on the detector interaction time~\footnote{To observe the Gaussian scaling it is more suitable to reformulate the $\mathrm{erfc}$-terms by means of Faddeeva functions.}. 
Thus, we find in Fig.~\ref{Fig1}(a) at $R/\sigma \approx 7.5$ a turning point in the timelike regime,  where the negativity constantly decays. At this point non-local correlations $\mathcal{M}(t_{BA}/T = 3)$ in Fig.~\ref{Fig1Ext}(a) drop below the local correlations and become constant. 
Before the turning point the $\mathrm{erfc}$-terms undergo a Gaussian decay. For smaller $t_{BA}/T$ the width of the Gaussian increases, shifting the turning point to larger $R/\sigma$.
Thus, for $t_{BA}/T = 0.5$ chosen in Figs.~\ref{Fig1}(b) and~\ref{Fig1Ext}(a) the width of the Gaussian has already increased so far that non-local correlations and negativity appear invariant under variation of $R/\sigma$.\\
When comparing Fig.~\ref{Fig1}(a.1) with Fig.\ref{Fig1}(a.2) we find, due to the additional factor of $\lambda_{\bm j}$ in local correlations, a revival of entanglement in the timelike regime after the turning point at $R/\sigma \approx 7.5$ is reached. This is caused by the narrowing of the frequency window in local correlations when increasing $\Omega T$. The overall lower entanglement amplitudes when comparing Figs.~\ref{Fig1}(a) with Figs.~\ref{Fig1}(b) is caused by the global Gaussian amplitude in detector energy in Eq.~\eqref{eq4.31}.
A deeper discussion on the overall influence of detector energy is provided in Sec~\ref{ImprSysPar}.
\\
In summary, the adiabatic nature of the switching function, the choice of Gaussian smearing and the axially symmetric setup are decisive for the dominant invariance under radial scaling of local correlations. With one exception for non-local correlations, where the choice of detector interaction time can have an amplifying effect as found in the case for small cavity radii in the micro and disc cavity regime.

\subsubsection{Imprint of cavity length}
\label{ICL}
    
We define, analogously to Eq.~\eqref{MLll}, neglecting all terms of Eq.~\eqref{eq4} not involved in the scaling of cavity length
    
\begin{subequations}
        \begin{align}
            L_{m} &=  \sum_l \frac{\sigma \,e^{-\left(\abs{\boldsymbol{k}_{\bm j}}\sigma\right)^2/2} }{L \,\abs{\boldsymbol{k}_{\bm j}}} e^{-\omega_{\bm j}^2T^2/2} \lambda_{\bm j} q_l 
            \label{LN},\\
            \begin{split}
            M_{m} &= \left| \sum_l \frac{\sigma \, e^{-\left(\abs{\boldsymbol{k}_{\bm j}}\sigma\right)^2/2} }{L \abs{\boldsymbol{k}_{\bm j}}} e^{-\omega_{\bm j}^2T^2/2}\right. \\
            &\hspace{.7cm}\times \left.\left( E(\omega_{\bm j}, t_{BA}) + E(\omega_{\bm j},-t_{BA}) \right) p_l \vphantom{\frac{e^{x^2}}{x^2}} \right|,
            \end{split}
            \label{MNnn}
        \end{align}
        \end{subequations}
        where we identified harmonic \mbox{$(\cos(k_l D))$} and parity \mbox{$((-1)^l)$} terms coupling to the longitudinal mode number
        \begin{subequations}
                    \begin{align}
            \begin{split}
             q_l 
             &=            \cos(\gamma_l^{(-)}) \cos(\gamma_l^{(-)})\\
             &= \frac{1}{2} \left[ 1 + (-1)^l \cos (k_l D)\right]\label{Par2},
             \end{split}
        \end{align}
       for the local correlations and
         \begin{align}
        \begin{split}
            p_l  
            &= \cos(\gamma_l^{(-)}) \cos(\gamma_l^{(+)})\\
            &= \frac{1}{2} \left[ \cos (k_l D) + (-1)^l \right],
            \end{split}
            \label{Par1}
            \end{align}
        \label{Par}%
\end{subequations}  
        
for the non-local correlations. 
The transversal modes contributing to the scaling of negativity in $L/\sigma$ are again selected by a Gaussian window. Combined with the prefactors of Eq.~\eqref{MNnn} emerging from Eqs.~\eqref{Eq.A} and~\eqref{Eq.Xi}, which scale with \mbox{$(L | \bm k_{\bm j}|)^{-1}$}, asymptotically a maximum for small cavity lengths occurs. This is followed by a decay when going to larger cavity lengths, resembling the asymptotic behavior of the negativity in Fig.~\ref{Fig1}.

For large enough cavities, the dominating terms of Eq.~\eqref{eq4}\footnote{Asymptotically we use the  approximation $\chi_{m}/R \sim m \pi/R$ which proves with an error of \mbox{$\sim (4 m)^{-1}$, cf. [52](9.5.12)} approximately similar in , sufficient for our task.}
scale approximately similar in $R/\sigma$ and $L/\sigma$.
Thus, we find when investigating non-local correlations for \mbox{$t_{BA}/T = 0.5$} in Fig.~\ref{Fig1Ext}(b), a decay dominated by the amplitude \mbox{$(L | \bm k_{\bm j}|)^{-1}$} and the Gaussian mode window with the \mbox{$E(\omega_{\bm j}, t_{BA}) + E(\omega_{\bm j},-t_{BA})$} terms appearing again as a constant amplitude.\\ 
For the choice of \mbox{$t_{BA}/T = 3$}, Fig.~\ref{Fig1Ext}(b) shows, equivalently to Fig.~\ref{Fig1Ext}(a), a strong decay in non-local correlations followed by a regime mostly invariant of the respective $x$-axis scaling. This invariant regime is the regime beyond the turning point of $L/\sigma \approx 7.5$. Nevertheless, the effect of the turning point is weakened when scaling $L/\sigma$ by still having a small slope, when compared to the $R/\sigma$ scaling, after the turning point is passed. This is caused by the different amplitude of \mbox{$(L | \bm k_{\bm j}|)^{-1}$} compared to the constant amplitude when scaling $R/\sigma$ beyond the turning point.
Due to the lack of \mbox{$E(\omega_{\bm j}, t_{BA}) + E(\omega_{\bm j},-t_{BA})$} in local correlations the $L/\sigma$ scaling of $\mathcal{L}$ in Fig.~\ref{Fig1Ext}(b) is fully dominated by the amplitude \mbox{$(L | \bm k_{\bm j}|)^{-1}$}, the Gaussian window and effects of the harmonic and parity terms.\\
Note that the harmonic terms in Eq.~\eqref{Par} undergo, by the choice of the smallest possible detector separation distance of \mbox{$D/\sigma = 5$}, strong averaging already for comparably small $L/\sigma$. 
This yields the parity term in Eq.~\eqref{Par1} of the local correlations averaging out with the harmonic term while the parity term in Eq.~\eqref{Par2} of the non-local correlations is conserved. In the setup at hand this has two reasons. First, the symmetric alignment of the detector system with respect to the cavity axis and the cavity mirrors at $z = 0$ and $z = L$. Second, the difference between local correlations, where the field is evaluated at each detector separately, and the non-local correlations, where the field is evaluated at different positions in the cavity. While for the local correlations the noise accumulated on each detector adds up constructively for opposite parity, the non-local correlations between the two detectors adds up destructively as it is shown in App.\ref{App3}~\footnote{Note that for the local correlations this is, as far as the single detectors do not influence each other by having overlapping wave functions, true not only for two but for an arbitrary number of detectors of arbitrary positioning in the cavity. For the non-local correlations this argument relies heavily on having only two detectors and their axial symmetric alignment.}. Thus, while each longitudinal mode interferes constructively to the local correlations, the longitudinal modes in the non-local correlations interfere destructively yielding a constant negativity already for mediate detector lengths of $L/\sigma > 2D$ as shown in Fig.~\ref{Fig1}(a).  
Additionally, if instead only on parity in Eq.~\eqref{MNnn} is selected, the two functions $p_l$ and $q_l$ have the same effect on local and non-local correlations yielding also to constructive interference of the longitudinal modes for the non-local correlations, thus boosting the negativity.

\begin{figure*}
\includegraphics[width=17.5cm]{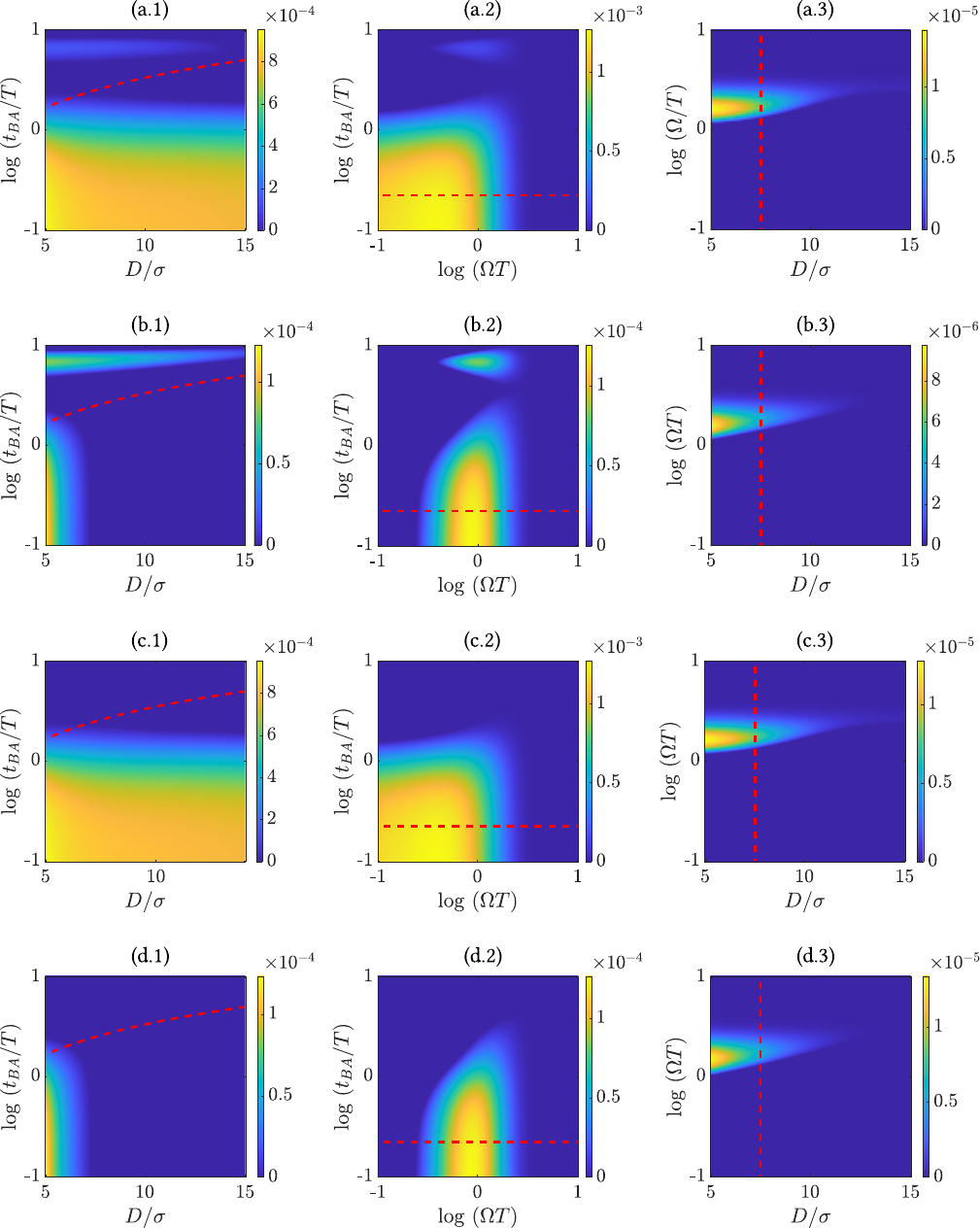}
\caption{Negativity estimator $\mathcal{N}$ for the four different cavity regimes. Each row resembles a different cavity regime: micro cavity ($L/\sigma = 20,\,R/\sigma = 10$) in (a), waveguide ($L/\sigma = 10^3,\,R/\sigma = 10$) in (b), disc cavity ($L/\sigma = 20,\,R/\sigma = 500$) in (c) and a optical cavity ($L/\sigma = 10^3,\,R/\sigma = 500$) in (d). In the first column the dimensionless equivalent of the detector distance $D/\sigma$ is varied with respect to the dimensionless separation time $t_{BA}/T$ of the detectors. Here we choose for the detectors energy gap $\Omega/T = 1$. In the second column $t_{BA}/T$ is varied for different $\Omega/T$. To obtain maximal negativity we choose the minimal detector separation of $D/\sigma = 5$. In the third column different $D/\sigma$ for different $\Omega/T$ are investigated for $t_{BA}/T = 2.5$. The red line marks the separation of timelike and spacelike regime.} 
\label{Fig2}
\end{figure*}

\subsection{Imprint of the detector system parameters}
\label{ImprSysPar}
        To bridge the gap between the preceding section~\ref{ImprintCavGeom} we examine each set of detector system parameters in each of the four different cavity regimes. These regimes include the micro cavity regime, the waveguide regime, the disc cavity regime and the optical cavity regime. All four different regimes with the specific scale of cavity parameters are shown in Fig.~\ref{FigReg}.\\  
        On top of that, and before starting the investigation of the detectors system parameters, a few points about the scales of the remaining variables have to be made.
        We choose for the range of detector separation distances \mbox{$D/\sigma \in [5,15]$}. For smaller detector distances we run into noise effects due to the overlap of the two detector wave functions, while for $D/\sigma = 5$ the overlap magnitude \mbox{$\exp(-\min (D/\sigma)^2/4) \approx 1.9 \times 10^{-3}$} is still neglectable. Beyond \mbox{$D/\sigma = 15$} a too large overlap of detector wave functions and cavity walls in the waveguide and disc cavity regime occurs. For the waveguide and the optical cavity the fast decay of the correlations when increasing the detectors separation distance makes $\max(D/\sigma) = 15$ again a good compromise.\\ 
        Additionally, for both $\Omega T$ and $t_{BA}/T$ we choose logarithmic scales from $[0.1,10]$ to account for larger ranges of detector energy and detector separation time. Increasing these ranges in both variables in Fig.~\ref{Fig2} beyond the chosen limits does not yield additional features. Thus, the chosen intervals cover the whole physics of both parameters. 
        The separation of timelike and spacelike regime are given by the dashed red lines in Figs.~\ref{Fig2} and~\ref{Fig3}. Note that in general for Gaussian switching the red lines also appear Gaussian~\cite{pozas2015harvesting}. However, due to the large scale of $t_{BA}/T$ and $D/\sigma$ we choose, without loss of generality, a hard cut for the edges of the light cone.
        When comparing microcavity with disc cavity regime and waveguide with optical cavity regime in Fig.~\ref{Fig2} we find the effect of increasing and decreasing the transversal mode density, despite only a few features, diminished.
        This is generated by the parity terms being not accessible for transversal modes due to the radial symmetric orientation of the detector system. The few features sensitive to the transversal mode spacing are all in the timelike regime as shown in \mbox{Fig.~\ref{Fig2}(a.1),(a.2),(b.1) and~(b.2)} and yield, compared to the maximum entanglement possible, only small entanglement amplitudes. It is also notable that for the system under investigation the largest amplitudes for entanglement harvesting occur, for all the different parameter regimes under investigation, solely in the spacelike regime.  

        \subsubsection{Detector separation time and detector separation distance}
        \label{secTD}
\begin{figure*}
    \includegraphics[width=17.5cm]{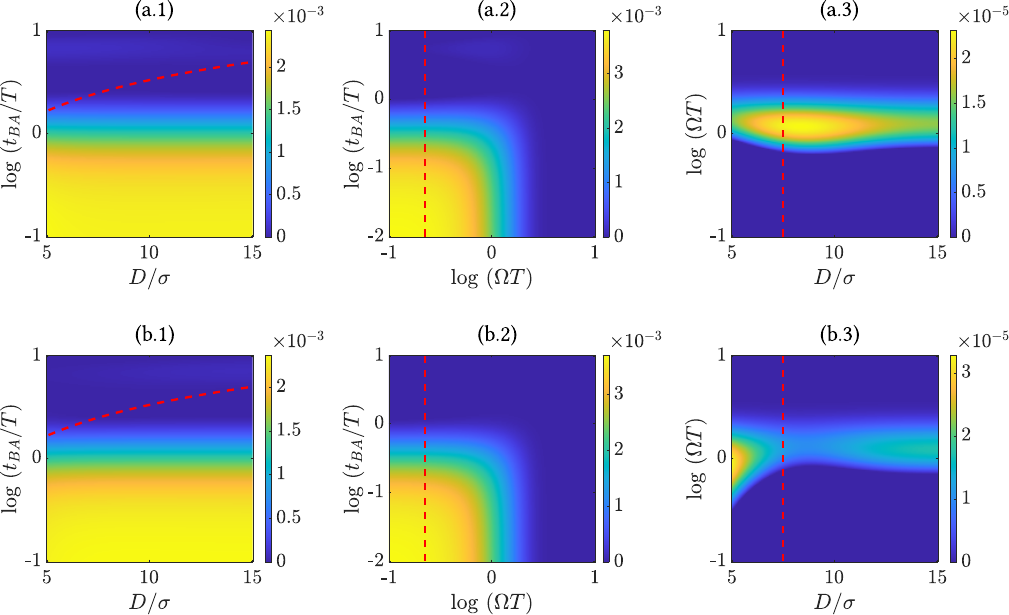}
    \caption{Negativity estimator $\mathcal{N}$ for the waveguide ($L/\sigma = 10^3,\,R/\sigma = 10$) with fixed parity in $l$. The numerical quantities $\Omega T$, $D/\sigma$ and $t_{BA}/T$
    in each row are chosen identical to Fig.~\ref{Fig2}. Even parity is shown in the first row (a) and odd parity is shown in the second row (b).} 
    \label{Fig3}
\end{figure*}

        For a deeper investigation we first consider the negativity dependence on the detector separation time $t_{BA}/T$ and the detector separation distance $D/\sigma$ in the first column of Fig.~\ref{Fig2}. We find for all four different cavity regimes a maximum in non-local correlations at the smallest detector separation and at small detector separation times which is the spacelike regime of \mbox{$D/\sigma = 5$} and \mbox{$t_{BA}/T < 0.5$}.\\ 
        When increasing \mbox{$D/\sigma$}, we find two different cases. 
        First, for large cavity lengths, i.e., for \mbox{$L/\sigma = 10^3$} in \mbox{Figs.~\ref{Fig2}(b.1) and~(d.1)}, there is a fast decay in negativity when increasing the detector separation distance. Separating the cosine expressions in harmonic and parity dependent expressions, as in Eq.~\eqref{Par}, we find that for long cavities the harmonic terms average out.  
        This is, the local correlations average out their $D/\sigma$ dependence, due to the large window of possible longitudinal mode numbers, as already discussed in Sec.~\ref{ICL}. The non-local correlations turn instead for the narrow longitudinal mode spacing into a cosine transform of the Gaussian low pass which starts decaying if \mbox{$D/\sigma$} reaches the correlation length of \mbox{$\mathcal{O}(\sigma)$}. If this scale is passed the local correlations surpass the non-local correlations which average to zero.\\
        Second, for small cavity lengths, i.e., for \mbox{$L/\sigma = 20$} in \mbox{Figs.~\ref{Fig2}(a.1) and~(c.1)}, only a small decay over the whole range of detector separation distance is visible. Due to the small length of the cavity only a few of the longitudinal modes have non neglectable contribution, not enough for the harmonic terms in Eq.~\eqref{Par} to average out. Thus, the correlation length is shifted to a larger order, yielding only a small decay when increasing the detectors interaction distance.\\
        Therefore, the denser the longitudinal mode spectrum, the smaller the detector separation distance can be chosen to still see significant entanglement between the two parties. On the one hand, this is caused by the noise given by the local correlations outgrowing the non-local correlations significantly for increasing $L/\sigma$ due to the different parity and harmonic functions of Eq.~\eqref{Par}. On the other hand, the harmonic terms can only contribute significantly if they interfere constructively with each other. This is, the larger the detector separation distance, the more phase incoherent the harmonic terms become.\\
        The separation in parity and harmonic terms, yields the harmonic terms as the Fourier coefficients of a discrete Fourier transform with respect to the detector separation distance. This is equivalent to the free space setup in~\cite{pozas2015harvesting}. Here the detector separation distance similarly appears as the phase of the Fourier transform of the non-local correlations and thus undergoes increasing averaging effects for increasing detector separation distance. Note that there -- as generally for detectors in free space -- the local correlations significant for the negativity $\mathcal{L}_{\mu \mu}$ do not depend on the detector separation distance. This translation invariance is broken in the cavity setup in general, but is restored for the limit of large cavity lengths. In this limit, as we find it with the harmonic term in Eq.~\eqref{Par1} averaging to zero if enough longitudinal modes contribute to the non-local correlations, translation invariance is restored.\\
        For different detector interaction times we find a decay in entanglement with increasing $D/\sigma$ in both, small and large cavity lengths, which does speed up with increasing \mbox{$t_{BA}/T$} until non-local correlations reach a constant plateau at \mbox{$t_{BA}/T = 3$}. Since at this plateau most of the non-local correlations have already decayed, entanglement also decays to zero when correlations converge towards the inside of the light cone by further increasing \mbox{$t_{BA}/T$}. This is especially made clear when comparing the first column of Fig.~\ref{Fig2}, where $t_{BA}/T$ is varied, with the third column of Fig.~\ref{Fig2}, where we choose $t_{BA}/T = 2.5$.\\
        When going inside the light cone we find a beam of width \mbox{$\Delta t_{BA}/T \approx 1.2$} centered at \mbox{$t_{BA}/T \approx 6.9$} in \mbox{Figs.~\ref{Fig2}(a.1) and~(b.1)}, where after decaying, entanglement harvesting becomes possible again. While the maximum of negativity starts to decay when increasing the cavity length, i.e., going from the micro cavity in the first row to the waveguide in the second row of Fig.~\ref{Fig2} the beam decays on a much slower rate, staying roughly constant compared to the decay of entanglement inside the light cone as seen in \mbox{Figs.~\ref{Fig2}(b.1) and~(d.1)}. When increasing the cavity radius, i.e., going to the disc cavity in the third row or the optical cavity in the fourth row, the beam vanishes instead. Thus, the influence of the radial mode spacing and thus of the cavities radius on the beam is much stronger than the influence of the longitudinal mode spacing controlled by the cavities length.
        For larger $t_{BA}/T$ where the beam occurs, the complementary error functions in Eq.~\eqref{Eq.E} have reached the constant plateau and thus are saturated, making the plane waves of $e^{\pm \omega_{\bm j} t_{BA}}$ dominating the shape of the function. If there is only a small number of radial field modes involved, as it is for the microcavity or the waveguide, the amplitude oscillates for fixed $l$ with a period of \mbox{$\Delta_l \approx 2\pi/(\omega_{(0,2,l)}-\omega_{(0,1,l)})$}. For the maximal longitudinal mode amplitude, given by $l = 0$, this yields $\Delta_0/T \approx 6.8$, which is the center of the beam described above.
        Thus, despite the large time delay between the interaction of the field on each of the two detectors, when in the regime of the beam, entanglement harvesting is possible again. The underlying mechanism here is governed by having only a small number of transversal field modes contributing to the detector-field interaction, as it is in the case of the micro cavity and the waveguide. 
        In these resonator regimes the Gaussian wave functions select only a few transversal field modes mediating the whole interaction of the transverse modes and the detectors.
        When in these resonator regimes the few dominating field modes add up coherently, entanglement can be conserved, despite an increased time interval between the interactions of the field with the detectors.
        When increasing the cavity radius we find that an averaging effect over a large number of transversal modes occurs, blurring out the sharp oscillations in the detector separation time and thus the beam for the case of disc cavity and the optical cavity in \mbox{Figs.~\ref{Fig2}(c.1) and~(d.1)}.\\
        Especially for the disc cavity, only a few longitudinal modes contribute and thus a similar effect as for the waveguide should be observed but with the \mbox{$\omega_{(0,1,0)}$} and \mbox{$\omega_{(0,1,1)}$} modes as carrier frequencies. Defining analogously to $\Delta_l$ a measure $\Delta_m$ where we hold $m$ fixed and choose $l = 0$ and $l =1$, we find \mbox{$\Delta_1/T \approx 13.3$}. Here the first beating occurs where the Gaussian switching has decayed to a level where no oscillations are visible anymore, making the revival of entanglement in the timelike regime of the simulated setup unique to the micro cavity and the waveguide.\\
        When splitting the longitudinal mode numbers in even $l$ and odd $l$, as it is shown in the first column of Fig.~\ref{Fig3} for the waveguide, the beam vanishes. Thus we can follow that, in contrast to the entanglement amplitude in the space like regime, the entanglement amplitude on the beam is invariant under parity splitting. 
        This is in contrast to the entanglement outside the light cone where opposite parities interfere destructively, causing a strong decline of non-local correlations for $D/\sigma >5$ as we find it in \mbox{Fig.~\ref{Fig2}(b.1)}.\\
        Therefore, the interaction time dependent amplitude~\eqref{Eq.E} in \mbox{Fig.~\ref{Fig2}(b.1)} acts with its monotonically decay for \mbox{$t_{BA}/T < 3$} like a filter weighting odd and even modes such that destructive interference occurs between odd and even parity. This is also seen for the case of the microcavity and the disc cavity, where due to their small length, no averaging effect of the parity term can occur, diminishing the strong decay for $D/\sigma > 5$. Conclusively, the entanglement outside the lightcone, decays much smoother and thus can exist for larger $D/\sigma$,while inside the lightcone the opposite is the case.

\subsubsection{Detector separation time and detector energy}
        In the second column of Fig~\ref{Fig2} the correlations dependence on the detector interaction time \mbox{$t_{BA}/T$} and the detector energy \mbox{$\Omega T$} is investigated.\\
        For both, the detector interaction time and the detector energy, we find again the Gaussian switching in all four different cavity regimes, when going to \mbox{$\Omega T>1$} and $t_{BA}/T>1$, respectively. For \mbox{$\Delta_0/T \approx 6.8$} the familiar beam occurs, which is now in shape of an arrowhead pointing towards the negative \mbox{$\Omega T$}-axis.\\ 
        The specific shape of the area with maximal entanglement accumulating around \mbox{$\Omega T = 1$} is caused by the coupling~\eqref{Coupllambda} in the local correlations. Here, $\Omega T$ couples directly to the field frequencies and thus the parity of each mode and does not solely occur as prefactor as we have it in the negativity~\eqref{eq4.31} for the remaining $\Omega T$ dependence.  When the parity terms average out for large cavity lengths $q_l$ in Eq.~\eqref{Par1} is unable to outgrow $p_l$ in Eq.~\eqref{Par2} for small $\Omega T$ where the coupling~\eqref{Coupllambda} maximizes with \mbox{$\lambda_{\bm j} \approx 1$}.\\
        This is also visible in the second column of Fig.~\ref{Fig3} where for constant parity in the waveguide regime the regions where entanglement harvesting is allowed does not decay for small $\Omega T$ and thus is of similar shape than for the micro cavity in \mbox{Fig.~\ref{Fig2}(a.2)}. One exclusion here is the arrowhead region which vanishes for constant parity. This is similar to the observations made when comparing the first column of Fig.~\ref{Fig2} with the first column of Fig.~\ref{Fig3}, which again confirms the time-dependent amplitude~\eqref{Eq.E} as a kind of parity filter.\\
        Additionally we find an unsymmetrical shape of non-local correlations around \mbox{$\Omega T = t_{BA}/T = 1$} and a preferred direction of non-local correlations dominating the local correlations when increasing both $\Omega T$ and $t_{BA}/T$ to larger $\Omega T$ and larger $t_{BA}/T$. This effect is increased in case of the waveguide and the optical cavity in \mbox{Figs.~\ref{Fig2}(b.2) and~(d.2)}, where for the former one we also find the above described arrowhead with the opening to the right. To investigate these features we define on these edges the detector frequency $\widetilde{\Omega T}$ where the entanglement vanishes. This is
        \begin{align}
            \mathcal{L}_{\alpha \alpha} &\leq \tilde{\mathcal{L}} =  e^{- (\Omega T)^2/2} e^{- \omega_{1,0} \Omega T^2} C,\\
            |\mathcal{M}| &= e^{-(\Omega T)^2/2} A(t_{BA}/T),
        \end{align}
        where $\tilde{\mathcal{L}}$ is the upper bound of $\mathcal{L}_{\alpha \alpha}$ when maximizing $\lambda_{\bm j}$ of Eq.~\eqref{Coupllambda} in $\omega_{\bm j}$ and $C$ and $A(t_{BA}/T)$ are real and positive amplitudes. With these equations we find 
        \begin{align}
            \widetilde{\Omega T} \propto \ln (C) - \ln (A(t_{BA}/T)).
        \end{align}
        For small longitudinal mode spacing as it is for the waveguide or the optical cavity a dense spectrum of field resonances occurs. In this case, the longitudinal sum can be transformed into an integral. Asymptotic expansion of $\omega_{\bm j}$ around $l = 0$ and phase matching yields a stationary wave number 
        \begin{align}
            \kappa_ {m} = \frac{2 D T}{ \sigma \,t_{BA}} k_{m,0}. 
        \end{align}
        Thus, for large cavity lengths the field frequency can continuously adjust to the new resonance of the detector by a change of interaction time, producing a sharp variation of \mbox{$A(t_{BA}/T)$} in $t_{BA}/T$. In contrast to that, the coarser spectrum and the lack of a zero mode in $m$ causes no increase in variation when increasing the cavity radius, as seen when comparing the waveguide in \mbox{Fig.~\ref{Fig2}(b.2)} to the optical cavity in \mbox{Fig.~\ref{Fig2}(d.2)}. This is also why the entanglement found for the microcavity in \mbox{Fig.~\ref{Fig2}(a.2)} and the disc cavity \mbox{Fig.~\ref{Fig2}(c.2)} seem, despite the arrowhead, identical. In these regimes, the smaller variation in $A(t_{BA}/T)$ is caused due to the coarser spectrum of the few $l$'s and thus the few field frequencies which are not suppressed by the Gaussian window. So as $(t_{BA}/T)$ changes, the field cannot adjust to the new stationary point. If adjustment happens a jump of the correlations is smoothed out by the Gaussian, yielding only the small slope observed in \mbox{Fig.~\ref{Fig2}(a.2) and~(c.2)}.\\
        If now only one parity is selected as we find in the second column of Fig.~\ref{Fig3} the sharp variation of \mbox{$A(t_{BA}/T)$}. We find that without parity cancellation the amplitude \mbox{$A(t_{BA}/T)$} looses its ability to adjust to the change of detector interaction time and a similar pattern in entanglement as for the micro cavity and the disc cavity in \mbox{Fig.~\ref{Fig2}(b.1) and~(b.3)} is reproduced.
        This results in not only larger entanglement amplitudes, but also possible entanglement for a larger regime of detector energies when applying parity selection in the case of the \mbox{waveguide}.\\
        On top of that we find for increasing $t_{BA}/T$ the complementary error function in amplitude~\eqref{Eq.E} decreasing. This relaxes the upper bound $\tilde {\mathcal{L}}_{\alpha \alpha}$, allowing larger detector energies, until the dominating Gaussian in \mbox{$\Omega T$} lets both local and non-local correlations drop.
        
    \subsubsection{Detector energy and detector separation distance}
    
            In the third column in Fig.~\ref{Fig2}, when investigating the detector energy $\Omega T$ and the detector separation distance $D/\sigma$, we find non-local correlations mostly dominating in the spacelike regime. The maximum is for all cavity regimes at detector energies between \mbox{$\Omega T = 1$} and \mbox{$\Omega T = 4$}. This is due to the chosen detector interaction time of \mbox{$t_{BA}/T = 2.5$}, where \mbox{$\widetilde{\Omega T} \rightarrow 1$} and is in accordance with the second column of Fig.~\ref{Fig2}.\\ 
            When increasing $D/\sigma$ non-local correlations start decaying stronger in the case of the waveguide and the optical cavity in \mbox{Figs.~\ref{Fig2}(b.3) and~(d.3)} compared to the micro cavity and the disc cavity in \mbox{Figs.~\ref{Fig2}(a.3) and~(c.3)}. This hierarchy in decay was already observed when analyzing the first column of Fig.~\ref{Fig2}. \\
            Nevertheless when investigating constant parity as in the third column of Fig.~\ref{Fig3} the main accumulation of entanglement seen for waveguide and optical cavity does not converge towards the pattern of micro cavity and disc cavity as observed in the subsequent discussions.\\
            To be more precise, we find that for even parity, cf. Fig.~\ref{Fig3}, the maximum of entanglement moves to the right. This is due to the cosine in Eq.~\eqref{Par} of the odd parity being shifted by a factor of \mbox{$\pi D/L$} relatively to the even parity. Therefore, the maxima occur shifted with the maximum of the even parity in Fig.~\ref{Fig3}(a) towards the interior of the lightcone, while for the odd parity in Fig.~\ref{Fig3}~(b.3) a shift towards the outside of the lightcone occurs. Another maximum of the cosine can be seen for the case of odd parity again for larger $l$ but with an amplitude suppressed by the Gaussian switching, which is the weak maximum seen around \mbox{$D/\sigma = 15$} in Fig.~\ref{Fig3}(b.3).\\
            Around the shifted maxima we also see a broadening in the $\Omega T$ direction. Due to the parity selection, the non-local correlations get concentrated around a specific maximum, yielding the upper bound $\tilde{\mathcal{L}}_{\alpha \alpha}$ allowing for a larger range $\Omega T$. 

\section{Conclusion}

        In this paper we derived the analytic expression for the negativity of two identical detectors, each with one ground and one excited state coupling to the vacuum correlations of a general cavity field for arbitrary switching and arbitrary detector wave functions.  
        On this basis we generalized for two Gaussian detectors placed mirror symmetric on the symmetry axis of a cylindrical cavity and chose Gaussian switching for the light matter coupling.  
        On top of that we presented a detailed asymptotic discussion focusing on multiple aspects influencing the observation and change in negativity due to variations in the cavities length scales. These aspects were, the change of local and non-local correlations, detector energy, detector separation distance, detector separation time and last but not least the influence of the fields parity. All investigations focused on both, the timelike and the spacelike regime.\\
        This shed light into the scaling of correlations when increasing cavity length and/or cavity radius which were found to be fundamentally different when comparing time and spacelike regime. 
        That is we identified regimes of the detector system parameters which seem unaffected when scaling the cavity radius and regimes unaffected when scaling the cavity length. 
        Including, on the one hand, when scaling the cavity length, a strong decay in the spacelike regime while correlations in the timelike regime seem to be mostly unaffected.
        On the other hand we find when scaling the cavity radius, a strong decay in the timelike regime while correlations in the spacelike regime seem to be mostly unaffected.
        This contrasts sharply with the dominant scaling in the cavity volume as specified by the normalization of the mode functions and thus of light matter coupling in a cavity in general.\\
        On top of that we find that the strong influence of parity allows a regaining of correlations when applying parity selection, but which was in the setup here only observed in the spacelike regime. Additionally, parity selection allows not only a larger entanglement amplitude but also increases the regime of detector frequency and detector separation distance for entanglement harvesting. \\
        Especially due to the Gaussian switching and the harmonic oscillator detector wave functions we find the setup maximizing entanglement for detectors with minimal distance and minimal interaction time with the electromagnetic field.\\
        Conclusively the non trivial interaction between vacuum correlations confined by a cavity and the detectors parameters allows controlling of entanglement in regimes not applicable for free space. By the interplay between cavity settings and the degrees of freedom controlled by the detector setup a  wide range of different entanglement magnitudes can be engineered, laying the path for future experiments.    

\section*{Acknowledgements}
JS and NM would like to especially thank W. P. Schleich, R. Lopp, M. Efremov, A. Wolf and  J. Seiler for stimulating and helpful discussions.
The CAL III project is supported by the German Space Agency at the German Aerospace Center (Deutsche Raumfahrtagentur im Deutschen Zentrum für Luft- und Raumfahrt, DLR) with funds provided by the Federal Ministry for
Economic Affairs and Climate Action (Bundesministerium
für Wirtschaft und Klimaschutz, BMWK) due to an enactment of the German Bundestag under Grant No. 50WM2545B (CAL III).

\appendix

\section{The respective Detector Orientation}
\label{AppDet}
\renewcommand{\theequation}{A.\arabic{equation}}

We recall that the general solution of the Schrödinger equation of a central potential $V(r)$ can be written in spherical coordinates through the wave function $\Psi_{klm}(r,\vartheta,\phi)=R_{kl}(r)Y_{lm}(\vartheta,\phi)$ where the radial function $R_{kl}(r)$ solves the radial dependence of the chosen potential and $Y_{lm}(\vartheta,\phi)$ are the spherical harmonics which govern the angular dependence\,\footnote{Supplementary, the quantum numbers $(k,l,m)$ can take the values $k=0,2,4,6,...$, $l=0,1,2,3,...$ and $m=-l,-l+1,...,l$.}. Furthermore, we know that the spherical harmonics are also eigenfunctions of the orbital angular momentum operator $\hat{L}^2$ as well as of its $z$-component $\hat{L}_z$. The rotation of the angular momentum operators under changes of reference frame is characterized by the Wigner-D matrix, transforming the spherical harmonics of detector $B$ to detector $A$'s frame linearly via 
\begin{align}
    Y_{lm}^B(\vartheta,\phi)=\sum_{\mu=-l}^lY_{l\mu}^A(\vartheta,\phi)D_{\mu,m}^l(\psi,\vartheta,\varphi).
\end{align}
Here $D_{\mu,m}^l(\psi,\vartheta,\varphi)$ are the elements of the Wigner-D matrix~\cite{pozas2016entanglement}. To this end, we transform \eqref{eq4.3} into spherical coordinates
\begin{align}
  \Psi_e^A(\boldsymbol{r})=\Psi_{010}(r,\vartheta,\phi)=R_{01}(r)Y_{10}^A(\vartheta,\phi)  
  \label{ESA}
\end{align}
where the wave function depends on \mbox{$Y_{10}(\vartheta,\phi)$}. Consequently, with the aforementioned relation, the angular wave function of detector $B$ with respect to detector $A$'s reference frame results into
\begin{equation}
  \begin{aligned}
    Y_{10}^B(\vartheta,\phi)&=\sum_{\mu=-1}^1Y_{1\mu}^A(\vartheta,\phi)D_{\mu,0}^1(\psi,\vartheta,\varphi)\\
    &=-\frac{1}{2}\sqrt{\frac{3}{\pi}}\sin(\vartheta)\sin(\vartheta)\cos(\phi+\psi)\\
    &\hspace{.5cm}+\frac{1}{2}\sqrt{\frac{3}{\pi}}\cos(\vartheta)\cos(\vartheta).
    \end{aligned} 
\label{eq4.5}
\end{equation}
In a next step, we insert \eqref{eq4.5} into the solution 
\begin{align}
\Psi_e^B(\boldsymbol{x})=R_{01}(r)Y_{10}^B(\vartheta,\phi).
\end{align}
A transformation into cylindrical coordinates using both representations of the position vector\footnote{For cylindrical coordinates we define \mbox{$\boldsymbol{r} = (\rho\cos(\phi),\rho\sin(\phi),z)^T$} and for spherical coordinates we use \mbox{$\bm r = (r\cos(\phi)\sin(\vartheta),r\sin(\phi)\sin(\vartheta),r\cos(\vartheta))^T$}}, yields
\begin{equation}
    \Psi_e^B(\boldsymbol{r})=\Psi_{000}^B(\boldsymbol{r})\frac{\sqrt{2}}{\sigma}[\rho\mathrm{sin}(\vartheta)\mathrm{cos}(\phi+\psi)+z\mathrm{cos}(\vartheta)].
    \label{ESB}
\end{equation}
The detector wave functions of Eq.~\eqref{ESA} and Eq.~\eqref{ESB} then define the
spatial smearing functions of Eq.~\eqref{Smear}.

\section{Analytical Calculation of the Negativity}
\label{App1}
\renewcommand{\theequation}{B.\arabic{equation}}

Due to the angular symmetry of the $TE$-mode and the detector setup, located on the symmetry axis of the cavity, the polarization $\mu_1$ vanishes. Since we have only one polarization contributing which are the $TM$-modes, i.e., $\mu = \mu_2$ we will neglect the polarization index $\mu$ in the following. To decompose the Wightman Tensor we also predefine the electric field amplitudes in position representation 
\begin{align*}
    \mathcal{E}_{\bm j}^{(+)}(\bm x, t) &= \langle \bm j |\hat{\bm E} (\bm x, t) | 0 \rangle,\\
    \mathcal{E}^{(-)}_{\bm j}(\bm x, t) &=  \langle 0 |\hat{\bm E} (\bm x, t) | \bm j \rangle,
\end{align*}
where the subscript $(+)$ is associated with the creation operators amplitude $\hat{a}_{\bm j}^\dagger$ and the subscript $(-)$ is associated with the annihilation operators amplitude $\hat{a}_{\bm j}^{\vphantom{\dagger}}$, respectively. 

\subsection{Calculation of the Local Term}
\label{App1.1}
\renewcommand{\theequation}{A.1.\arabic{equation}}

The local term given explicitly in Eq.~\eqref{eq4.22} yields the local interaction with the field, i.e., the interaction of each detector with the field individually. Since we choose two detectors of equal smearing, the only difference between these two local interaction terms is the different position of the detectors. Since the local correlations of each detector are independent of the other detectors local correlations and vice versa, the respective orientation of both detectors to each other is omitted here. 
The structure of Eq.~\eqref{eq4.22}, allows to define a quantity $L_{\alpha\alpha, \bm j}$ such that
\begin{align}
    \mathcal{L}_{\alpha \alpha} = \sum_{\bm j} |L_{\alpha\alpha,\bm j}|^2,
    \label{App:AbsL}
\end{align}
with $\alpha \in \{A,B\}$ and the function
\begin{widetext}
\begin{align}
    L_{\alpha\alpha,\bm j} = &e\int_{-\infty}^{\infty}\mathrm{d}t_1\int\mathrm{d}^3 x_1\,\,e^{i\Omega t_1}\chi_{\alpha}(t_1) \mathcal{E}_{\bm j}^{(+)}  (\bm x_1,t_1) \cdot\boldsymbol{F}(\boldsymbol{x}_1-\boldsymbol{x}_{\alpha}),
\end{align}
reducing the number of integrals to calculate~\eqref{eq4.22} from eight to four. Note that we assumed real switchings $\chi_\alpha (t)$. 
For the detector $A$, located at $z = (L-D)/2$ and the detector $B$, located at $z = (L +D)/2$, we find
\begin{align}
    L_{AA,\bm j} = &A_{\bm j} \int_{-\infty}^{\infty}\mathrm{d}t_1\int\mathrm{d}^3 x_1\, e^{-\frac{(t_A - t_1)^2}{T^2}} e^{- \frac{r^2}{\sigma^2} - \frac{(2z +D - L)^2}{4\sigma^2}
- i m \phi} e^{ i t_1 (\omega_{\bm j} +  \Omega)}
\left[
k_{nm} L (2z - D - L)\, J_{m}\!\big(k_{nm} r\big)\,\cos(k_l z)\right. \nonumber\\
&\left.- k_l L\, r\,\Big(J_{m-1}\!\big(k_{nm} r\big) - J_{m+1}\!\big(k_{nm} r\big)\Big)\,\sin(k_l z)
\right](2z + D -L),
\label{App:LAExt}\\
L_{BB,\bm j} = &A_{\bm j} \int_{-\infty}^{\infty}\mathrm{d}t_1\int\mathrm{d}^3 x_1\, e^{-\frac{(t_B - t_1)^2}{T^2}} e^{- \frac{r^2}{\sigma^2} - \frac{(d - L + 2z)^2}{4\sigma^2}
- i m \phi} e^{ i t_1 (\omega_{\bm j} +  \Omega)}
\left[
k_{nm} L (d - L + 2z)\, J_{m}\!\big(k_{nm} r\big)\,\cos(k_l z)\right. \nonumber\\
&\left.- k_l L\, r\,\Big(J_{m-1}\!\big(k_{nm} r\big) - J_{m+1}\!\big(k_{nm} r\big)\Big)\,\sin(k_l z)
\right],
\label{App:LBExt}
\end{align}
with the mode number dependent amplitude
\begin{align}
    A_{\bm j} = \frac{e c}{2 \pi^2 \sigma^4 R \sqrt{2 \epsilon_0 L^3 \hbar \,\omega_{\bm j} J_{n+1}^2(k_{nm}R)}}.
\end{align}
For both detectors the $\varphi$-symmetry of the problem fixes the radial mode number $n = 0$, yielding only a prefactor of $2 \pi$ from the angluar integration. For the $z$-integration we use the strong localization of the detectors in cavity to extend the integration limits $z \in [0,L]$ to the whole space of $\mathbb{R}$. Under these assumptions the spatial integration of Eq.~\eqref{App:LAExt} and Eq.~\eqref{App:LBExt} reduce to integrals of the kind found in~\cite{SL}, App. G., yielding 
\begin{align}
L_{AA,0,m,l} = 
&B_{\bm j}e^{- \frac{k_{0,n,l}^2 \sigma^2}{4}}    \cos (\gamma_l^{(-)}) \int_{-\infty}^{\infty}\mathrm{d}t_1\,
e^{-\frac{(t_A - t_1)^2}{T^2}}
e^{i t_1 (\Omega
+\omega_{0,n,l})}\\
\label{App:LExt2}
L_{BB,0,m,l} = 
&B_{\bm j}e^{- \frac{k_{0,n,l}^2 \sigma^2}{4}}  \cos (\gamma_l^{(+)}) \int_{-\infty}^{\infty}\mathrm{d}t_1\,
e^{-\frac{(t_B - t_1)^2}{T^2}}
e^{i t_1 (\Omega
+\omega_{0,n,l})}
\end{align}
with the prefactor 
\begin{align}
    B_{\bm j} = 2  A_{\bm j} k_m \sqrt{\pi}^3 L \sigma^5. 
\end{align}
The time integration in Eq.~\eqref{App:LExt2} is a Gaussian integral 
\begin{align}
\int_{-\infty}^{\infty}\mathrm{d}t_1\,
e^{-\frac{(t_A - t_1)^2}{T^2}}
e^{i t_1 (\Omega
+\omega_{0,n,l})}
= \sqrt{\pi}\, T e^{i t_A (\Omega + \omega_{\bm j})} e^{-\frac{(\Omega + \omega_{\bm j})^2T^2}{4}} 
\label{App:LTint}
\end{align}
Due to the absolute value in~\eqref{App:AbsL} the phase term dependent on the detectors eigentime $t_A$ and $t_B$, respectively, in Eq.~\eqref{App:LTint} drop out. Additionally we find
\begin{align}
    \cos(\gamma_l^{(-)})^2 = \left[\cos(k_l L) \cos(k_l \frac{D}{L}) + \sin(k_l L) \sin(k_l \frac{D}{L}) \right]^2 = \cos(k_l \frac{D}{L})^2 =  \cos(\gamma_l^{(+)})^2.
\end{align}
These conditions yield 
\begin{align}
    \mathcal{L}_{AA} = \mathcal{L}_{BB} = B_{\bm j}^2 \,\pi \, T^2e^{-\frac{(\Omega + \omega_{\bm j})^2T^2}{2}} e^{- \frac{k_{0,n,l}^2 \sigma^2}{4}}    \cos^2 \left(\gamma_l^{(-)}\right),
\end{align}
which is equivalent to Eq.~\eqref{LExplicit}.

\subsection{Calculation of the non-local Term}
\label{App1.2}
\renewcommand{\theequation}{A.2.\arabic{equation}}

We can decompose the non-local correlations of Eq.~\eqref{eq4.23} as follows
\begin{align}
    \mathcal{M} = \mathcal{M}_{AB} + \mathcal{M}_{BA},
\end{align}
where we defined 
\begin{align}
    \mathcal{M}_{AB} &= -\sum_{\bm j} \int_{-\infty}^{\infty}\mathrm{d}t_1\int_{-\infty}^{t_1}\mathrm{d}t_2 \, g^{\vphantom{\ast}}_{\bm j}(t_A,t_1,\Omega) f^\ast_{\bm j} (t_B,t_2,-\Omega,\vartheta,\psi),
    \label{App:MAB}\\
     \mathcal{M}_{BA} &= -\sum_{\bm j}\int_{-\infty}^{\infty}\mathrm{d}t_1\int_{-\infty}^{t_1}\mathrm{d}t_2 \, g^\ast_{\bm j}(t_B,t_1,-\Omega) f^{\vphantom{\ast}}_{\bm j}(t_A,t_2,\Omega,-\vartheta,-\phi).
     \label{App:MBA}
\end{align}
In the last equation we already implied the change of the detector sequence from $AB$ to $BA$. When applying this operation not only the sequence of the switching functions has to be changed, but also the Euler angles have to be switched accordingly by \mbox{$\vartheta \rightarrow-\vartheta$} and \mbox{$\psi \rightarrow - \phi$}, (cf.~\cite{pozas2016entanglement}, App. C.). However, for the azimuthal symmetric setup considered, these changes will have no effect. 
Following Eq.~\eqref{eq4.24} we find
\begin{align}
    g_{\bm j}(t,t_1,\Omega) = -e\int\mathrm{d}^3 x \,e^{i\Omega t_1} \chi(t,t_1)\boldsymbol{F}^{T}(\boldsymbol{x}-\boldsymbol{x}_{A}) \cdot \mathcal{E}^{(-)}_{\bm j}(\bm x, t_1)
    \label{EqAppg},\\
    f_{\bm j}(t,t_2,\Omega;\alpha,\beta) =  -e \int\mathrm{d}^3 x \,e^{i\Omega t_2} \chi(t,t_2)\boldsymbol{F}_{\mathcal{R}}^{T}(\boldsymbol{x}-\boldsymbol{x}_{B};\alpha,\beta) \cdot \mathcal{E}^{(-)}_{\bm j}(\bm x, t_2),
    \label{EqAppf}
\end{align}
reducing the number of integrals found in Eq.~\eqref{eq4.24} from twelve to six.
Note that we here again assume real smearing functions.
Inserting the electric field~\eqref{EField}, switching function~\eqref{GS} and smearing functions~\eqref{Smear} into Eq.~\eqref{EqAppg} and Eq.~\eqref{EqAppf} yields 
\begin{align}
g_{\bm j} (t,t_1,\Omega) = &A_{\bm j}\,
e^{-\frac{(t - t_1)^2}{T^2}} e^{- \frac{r^2}{\sigma^2} - \frac{(d - L + 2z)^2}{4\sigma^2} + i m \phi} e^{- i t_1 \omega_{\bm j} + i t_1 \Omega}\,r\,(d - L + 2z)\nonumber\\
&\left[
k_{nm} L (d - L + 2z)\, J_{m}\!\big(k_{nm} r\big)\,\cos(k_l z) - k_l L\, r\,\Big(J_{m-1}\!\big(k_{nm} r\big) - J_{m+1}\!\big(k_{nm} r\big)\Big)\,\sin(k_l z)
\right],\label{gjexpl}\\
f_{\bm j}(t,t_2,\Omega;\alpha,\beta) = &
A_{\bm j}\,
 e^{-\frac{(t - t_2)^2}{T^2}} e^{- \frac{r^2}{\sigma^2} - \frac{(d + L - 2z)^2}{4\sigma^2}
+ i m \phi} e^{ - i t_2 \omega_{\bm j} + i t_2 \Omega}\,r \left[(d + L - 2z)\cos(\alpha) + 2r\cos(\phi+\beta)\sin(\alpha)\right]\nonumber\\
&\left[
k_{nm} L (d + L - 2z)\, J_{m}\!\big(k_{nm} r\big)\,\cos(k_l z)
+ k_l L\, r\,\Big(J_{m-1}\!\big(k_{nm} r\big) - J_{m+1}\!\big(k_{nm} r\big)\Big)\,\sin(k_l z)
\right]\label{fjexpl},
\end{align}
Again, when integrating over the angle $\phi \in[0,2\pi]$ we find that the angular mode number $n$ in Eq.~\eqref{gjexpl} gives only non vanishing couplings for $n = 0$, reducing the $\varphi$ integrals to a factor of $2 \pi$. For the $z$-integration we make again use of the strong localized detectors to increase the integration limits to the whole space of $\mathbb{R}$. Using this assumption the longitudinal and radial integration reduce again to integrals of the kind found in~\cite{SL}, App. G. yielding,
\begin{align}
g_{\bm j}(t,t_1,\Omega) = 
&B_{\bm j}\,
e^{-\frac{(t - t_1)^2}{T^2}}
e^{- k_{0,n,l}^2 \sigma^2 }
e^{i t_1 (\Omega
-  \omega_{0,n,l})}
 \cos (\gamma_l^{(-)})
 \label{gjcal},\\
f_{\bm j}(t,t_2,\Omega,\alpha,\beta) = &B_{\bm j}
\,e^{-\frac{(t - t_2)^2}{T^2}}
e^{- k_{0,n,l}^2 \sigma^2 }
e^{i t_2 (\Omega
-  \omega_{0,n,l})}
 \cos (\gamma_l^{(+)}) \cos (\alpha) 
 \label{fjcal}
\end{align}
For the time integration in Eq.~\eqref{App:MAB} we use the integral computed in~\cite{pozas2015harvesting}, App. A, and find
\begin{align}
\begin{split}
    &\hspace{0.8cm}\int_{-\infty}^{\infty}\mathrm{d}t_1\int_{-\infty}^{t_1}\mathrm{d}t_2 \,\,e^{-\frac{(t_A - t_1)^2}{T^2}}e^{i t_1 (\Omega- \omega_{0,n,l})}
    \,e^{-\frac{(t_B - t_2)^2}{T^2}}
    e^{i t_2 (\Omega+\omega_{0,n,l})}\\ 
    &= \frac{\pi T^2}{2} e^{-\frac{(\omega_{\bm j}^2 +\Omega^2)}{2}}e^{i t_A (\Omega + \omega_{0,n,l})}e^{i t_B (\Omega - \omega_{0,n,l})} \left[1 + i\,\mathrm{erf} \left( \frac{i \,(t_A - t_B) + T^2 \omega_{0,n,l}}{\sqrt{2}T} \right) \right].
\end{split}
\end{align}
The time integration in Eq.~\eqref{App:MBA} follows equivalently by applying complex conjugation and replacing $\Omega$ by $-\Omega$.
Combining this result with the spatial integrals of Eq.~\eqref{gjcal} and Eq.~\eqref{fjcal} we find the nonlocal term of the negativity presented in Eq.~\eqref{MExplicit}.

\section{Parity properties of local and non-local correlations}
\label{App3}
\renewcommand{\theequation}{C.\arabic{equation}}

Consider the two symmetrically aligned detectors on the cavity axis. The correlations measured at each detector are given by the function $\phi(z_i)$, with \mbox{$z_{1/2} = (L \mp d)/2$}. For the even/odd correlations we define
\begin{align}
    \phi_\pm (z ) = \frac{\phi(z) \pm \phi(L - z)}{\sqrt{2}}.
\end{align}
Thus, in terms of even and odd solutions we measure on each detector the correlations
\begin{subequations}
\begin{align}
    \phi(z_1) &=  \frac{\phi_+(z_1) + \phi_-(z_1)}{\sqrt{2}},\\
    \phi(z_2) &=  \frac{\phi_+(z_1) - \phi_-(z_1)}{\sqrt{2}}.
\end{align}
\end{subequations}
For the local correlations we find
\begin{subequations}
\begin{align}
    \mathcal{L} \sim \langle \phi^2(z_1)  \rangle + \langle \phi^2(z_2)   \rangle = \langle \phi_+^2(z_1)  \rangle + \langle \phi_-^2(z_1) \rangle
\end{align}
and for the non-local correlations 
\begin{align}
    \mathcal{M} \sim |\langle \phi (z_1) \phi (z_2)  \rangle | = | \langle \phi_+^2(z_1)  \rangle  - \langle \phi_-^2(z_1) \rangle |
\end{align}
\label{APPPAR}%
\end{subequations}
Note that Eqs.~\eqref{APPPAR} also hold for $\phi_{\pm} $ evaluated at $z_2$ instead of $z_1$ on the right hand side.
Thus, in the case of the axially symmetric setup the local correlations interfere correlations with different parity always constructively while the non-local correlations interfere correlations with different parities destructively.

\end{widetext}

\clearpage

\bibliography{Bib}

\end{document}